# Semantic-Sensitive Web Information Retrieval Model for HTML Documents

**Youssef Bassil**
*LACSC – Lebanese Association for Computational Sciences
Registered under No. 957, 2011, Beirut, Lebanon*
E-mail: youssef.bassil@lacsc.org

**Paul Semaan**
*LACSC – Lebanese Association for Computational Sciences
Registered under No. 957, 2011, Beirut, Lebanon*
E-mail: paul.semaan@lacsc.org

## Abstract

With the advent of the Internet, a new era of digital information exchange has begun. Currently, the Internet encompasses more than five billion online sites and this number is exponentially increasing every day. Fundamentally, Information Retrieval (IR) is the science and practice of storing documents and retrieving information from within these documents. Mathematically, IR systems are at the core based on a feature vector model coupled with a term weighting scheme that weights terms in a document according to their significance with respect to the context in which they appear. Practically, Vector Space Model (VSM), Term Frequency (TF), and Inverse Term Frequency (IDF) are among other long-established techniques employed in mainstream IR systems. However, present IR models only target generic-type text documents, in that, they do not consider specific formats of files such as HTML web documents. This paper proposes a new semantic-sensitive web information retrieval model for HTML documents. It consists of a vector model called SWVM and a weighting scheme called BTF-IDF, particularly designed to support the indexing and retrieval of HTML web documents. The chief advantage of the proposed model is that it assigns extra weights for terms that appear in certain pre-specified HTML tags that are correlated to the semantics of the document. Additionally, the model is semantic-sensitive as it generates synonyms for every term being indexed and later weights them appropriately to increase the likelihood of retrieving documents with similar context but different vocabulary terms. Experiments conducted, revealed a momentous enhancement in the precision of web IR systems and a radical increase in the number of relevant documents being retrieved. As further research, the proposed model is to be upgraded so as to support the indexing and retrieval of web images in multimedia-rich web documents.

## 1. Introduction

Information Retrieval (*IR*) is the science and practice of storing data, searching for data, and for information within data (Salton and McGill, 1983 ; Singhal, 2001). In a computing context, data are raw entities manageable by a computer device; they include, but not limited to text documents, web documents, images, graphs, videos, and audio clips. After several decades of effort, research, and development, the *IR* field has matured substantially and it became widely diverse and pervasive, as well as capable of retrieving textual and non-textual information out of millions of documents in just few seconds. The invention of the Internet in the late 60's and the World Wide Web in the early 90's, brought a necessity for users to search the web and locate online information in a consistent and efficient manner. Thus search engines were conceived. In essence, a web search engine such as Google, Yahoo, and Bing is an online *IR* system whose purpose is to grab web contents dispersed over the web and index them into searchable databases that allow Internet users to look up keywords and retrieve relevant web pages (Berry and Browne, 2005). Predominantly, most of the web pages available on the Internet are written using the Hyper Text Markup Language (*HTML*) which is composed of a set of markup tags that describe the content, the presentation, and the layout of the web page (Deitel, Deitel, and Nieto, 2002). Unfortunately, little work has been done to effectively model *HTML* documents in an information retrieval context, in such a way to associate the inner structure of the *HTML* language with the semantics of the document.

This paper proposes a Semantic-Sensitive Web Information Retrieval Model for *HTML* Documents that consists of a vector model called Semantic-Sensitive Web Vector Model (*SWVM*) and a term weighting scheme called Boosted Term Frequency-Inverse Document Frequency (*BTF-IDF*). *SWVM* is algebraically defined as $d = (\{w_{1,1}; w_{1,2}; w_{1,q}\}, \{w_{2,1}; w_{2,2}; w_{2,r}\}, \ldots, \{w_{i,1}; w_{i,2}; w_{i,s}\})$ where $i$ denotes a particular distinct term in $d$, $w_i$ denotes the weight of the $i_{th}$ term, and $w_{i,s}$ denotes the weight of the $s_{th}$ synonym for the $i_{th}$ term. It is semantic-sensitive because it stores, in addition to the term weight $w_i$, the weight of every synonym $w_{i,s}$ for every term $i$. Synonyms for terms are generated using a thesaurus during the indexing phase and are aimed to increase the likelihood of retrieving documents with similar context but different vocabulary terms. In order to weight terms i.e. calculating their number of occurrence in a particular document $d$, a new term weighting scheme is proposed called Boosted Term Frequency-Inverse Document Frequency (*BTF-IDF*). It is fundamentally based on calculating term weights using the classical Term Frequency-Inverse Document Frequency (*TF-IDF*), however, enhanced by boosting the weight of terms located within some special *HTML* tags. Particularly, the terms that are located within the tags *<title>*, *<meta>*, and *<h1>*, and terms that are part of the *URL* (Uniform Resource Locator) are assigned a higher weight than those located within other types of tags.

## 2. Background

In order for documents and queries to be manipulated by an *IR* system, they must first be transformed into a mathematical model often known as the vector space model (*VSM*) which contains several dimensions each corresponding to a particular term weight value. These values can be computed using different term weighting techniques and algorithms. In this section, the *VSM* and its best known weighting schemes are to be discussed elaborately, shedding the light on their advantages and disadvantages.

### 2.1. The Vector Space Model (VSM)

In modern information retrieval systems, terms or words that have occurred frequently in a text should have a higher score than the other terms (Salton and Buckley, 1988). This score is more specifically called term weight and indicates how much a term is significant with respect to its context (Salton, Wong, and Yang, 1975). Basically, a term weight denotes the frequency or the number of occurrence of a particular term in a document or in a collection of documents. A Vector Space Model (*VSM*) is an algebraic feature vector-based representation for documents and queries with their corresponding terms weight (Salton, Wong, and Yang, 1975). For instance, a document containing five terms *car*, *home*, *drive*, *garage*, and *summer* with term frequencies 10, 2, 4, 8, and 1 respectively, is represented using *VSM* as $d = (10,2,4,8,1)$. More generally, a feature vector for a document $d_j$ is represented as $d_j = (w_{1j},$ $w_{2j}$, $w_{3j}$, $\cdots$, $w_{nj}$) where $d_j$ denotes a particular document, $n$ denotes the number of distinct terms in document $d_j$, and $w_{ij}$ denotes the weight for the $i_{th}$ term in document $d_j$. Furthermore, in order to compare between documents or between documents and queries, i.e., determining the similarities between two documents $d_1$ and $d_2$, or between a document $d$ and a query $q$, the cosine of the angles between $d$ and $q$ is calculated. If the two documents are alike, they will receive a cosine of 1; if they are orthogonal (sharing no common terms) on the other hand, they will receive a cosine of 0 (Jurafsky and Martin, 2008). Formally, the similarity between document $d_j$ and query $q$ can be calculated using the following cosine equation:

$$sim(d_j, q) = \cos(\vec{d_j}, \vec{q}) = \frac{\vec{d_j} \bullet \vec{q}}{|\vec{d_j}||\vec{q}|} = \frac{\sum_{i=1}^{N} w_{i,q} \times w_{i,j}}{\sqrt{\sum_{i=1}^{N} w_{i,q}^2} \times \sqrt{\sum_{i=1}^{N} w_{i,j}^2}}$$

## 2.2. Existing Term Weighting Schemes

Different term weighting schemes exist, the best known are the Boolean weighting (Polettini, 2004), Term Frequency (*TF*) (Luhn, 1957), Logarithmic weighting (Kolda, 1997), Inverse Document Frequency (*IDF*) (Spark Jones, 1972), Term Frequency-Inverse Document Frequency (*TF-IDF*) (Salton and Buckley, 1988), and Entropy weighting (Dumais, 1991).

### 2.2.1. Boolean Term Weighing

It is also called binary weighing because it assigns binary weights for terms based on their presence or absence in the document (Polettini, 2004); a term is assigned a value of 1 if it appears at least one time in the document, and is assigned a value of 0 if it never appears. More generally, binary weights can be computed using the following equation:

$$\chi(t) = \begin{cases} 1 & \text{if } t > 0 \\ 0 & \text{if } t = 0 \end{cases}$$

Apparently, the Boolean method is not suitable for document weighting; rather, it more fits query weighting as it does not distinguish between terms that appear one time and terms that appear frequently.

### 2.2.2. Term Frequency (TF)

The raw frequency of a term within a document is called term frequency, and is defined as $tf_{t,d}$ with subscripts $t$ denoting a particular term and $d$ denoting a particular document (Luhn, 1957). Put simply, *TF* is the number of occurrence (frequency) of a particular term in a document. *TF* usually reflects the notion that terms occurring frequently in a document may be of greater importance than terms occurring less frequently, and therefore should have a heavier weight. Although *TF* is still considered to some extent an efficient approach to weight terms due to its simplicity and efficiency, it gives too much credit for terms that appear frequently and hence it discriminates less important documents over more important ones (Kolda, 1997).

### 2.2.3. Logarithmic Term Weighting

Logarithmic term weighting attempts to minimize the effect of frequency so that to balance the big variances in term weights and reduce the discrimination power of terms that appear too frequently (Kolda, 1997). The idea behind using *log* is that not always a term that appears five times is five times more important that the term that appears only one time.

### 2.2.4. Inverse Document Frequency (IDF)

The purpose of *IDF* is to attenuate the weight for terms that occur too often in a collection of documents and to increase the weight for terms that occur infrequently (Spark Jones, 1972). In other words, rare terms have high *IDF* and common terms have low *IDF*. For instance, a term such as "the" is likely to have a very low *IDF*, somewhat close to 0 because it would appear in every document; while, a term such as "psychophysics" is likely to have a very high *IDF* because it would appear in very few documents. *IDF* is generally defined as:

$$idf_j = \log\left(\frac{n}{df_j}\right)$$

$idf_j$ is the Inverse Document Frequency, $df_j$ is the number of documents in which term *j* occurs, and *n* is the total number of documents in the collection. Since the collection may encompass a huge number of documents, the value of $idf_j$ is usually reduced using the *log* function.

### 2.2.5. TF-IDF
Term Frequency-Inverse Document Frequency (*TF-IDF*) weights a particular term by calculating the product of its term frequency (*TF*) in a document with the *log* of its inverse document frequency (*IDF*) in the collection, and thereby discriminating terms that are frequent in a particular document but globally rare in the collection. In other words, terms that are frequent in a given document and infrequent in the whole collection are assigned high *TF-IDF* weight. *TF-IDF* is generally defined as:

$$w_{ij} = tf_{ij} \cdot idf_j$$

$w_{ij}$ is the final weight for term *j*, $tf_{ij}$ is the frequency of term *j* in document *i*, and $idf_j$ is the Inverse Document Frequency of term *j*. *TF-IDF* is by far the most successful document term weighting scheme and is applicable to almost all vector space information retrieval systems (Salton and Buckley, 1988).

### 2.2.6. Entropy Term Weighting
The Entropy weighting scheme assigns a weight of 0 for terms that appear once in every document, and a weight of 1 for terms that appears once in one document, and a weight between 0 and 1 for other combination of frequencies. The advantage of Entropy weighting is that it takes into consideration the distribution of terms over the collection by assigning higher weight for terms that occur less in a small number of documents (Dumais, 1991).

## 3. The Proposed Web Retrieval Model
The proposed web information retrieval model comprises two basic elements: A vector space model called *SWVM* short for Semantic-Sensitive Web Vector Model, and a term weighting scheme called *BTF-IDF* short for Boosted Term Frequency-Inverse Document Frequency. In essence, *SWVM* is based on the Vector Space Model (*VSM*) originally proposed by Salton, Wong, and Yang (1975), however, adjusted to exploit hypertext language in a way not to treat a web document as a regular classical unstructured text, but to take into account its tag structure and significantly boosts the weight of terms that appear in certain tags that are associated with the semantics of the document. For this purpose, *SWVM* employs a new term weighting scheme called *BTF-IDF* short for Boosted Term Frequency-Inverse Document Frequency, based on the classical Term Frequency-Inverse Document Frequency (*TF-IDF*), however, modified to assign a bonus weight for terms that occur in some specific *HTML* tags, namely <title>, <meta>, and <h1>, and for terms that are part of the document's *URL*. Characteristically, the proposed model is semantic-sensitive because it generates terms synonyms and integrates them as part of the vector's features so as to improve the precision-recall of the system, particularly when processing documents and queries having similar context but different vocabulary terms. Figure 1 depicts the document modeling phase whose function is to 1) fetch web documents, 2) parse *HTML* tags and extract terms, 3) weight these terms using the *BTF-IDF* weighting scheme, 4) convert the document into an *SWVM* model, and 5) index the final results into a relational database.

**Figure 1:** Document modeling phase

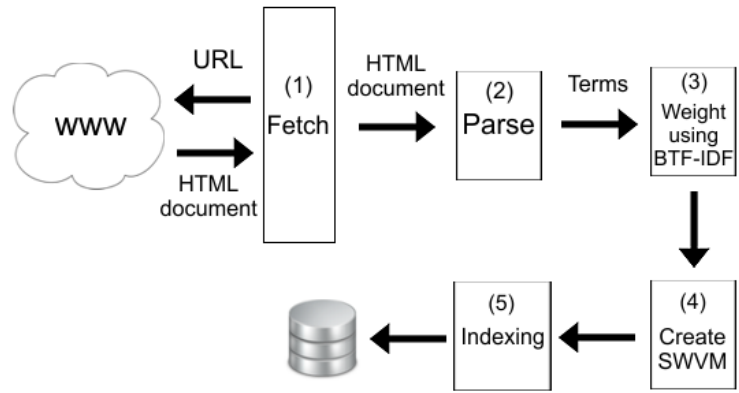

### 3.1. The SWVM Vector Model

*SWVM* can be mathematically defined as $d = (\{w_{1,1}; w_{1,2}; w_{1,q}\}, \{w_{2,1}; w_{2,2}; w_{2,r}\}, \ldots, \{w_{i,1}; w_{i,2}; w_{i,s}\})$ where $i$ denotes a particular distinct term in $d$, $w_i$ denotes the weight of the $i_{th}$ term, and $w_{i,s}$ denotes the weight of the $s_{th}$ synonym for the $i_{th}$ term. It is semantic-sensitive because it stores, in addition to the term weight $w_i$, the weight of every synonym $w_{i,s}$ for every term $i$. Synonyms are generated using a thesaurus during the indexing phase and stored together with other terms. The use of synonyms strengthens the retrieval system and allows a higher query-document match, especially when a query has similar context (similar meaning) with respect to a document but different vocabulary terms (Garcia, 2006). A given document $d$ that contains the terms "highest", "building", and "price" with term frequencies 3, 7, and 1 respectively, can be represented using *SWVM* as:

$d = (\{w(\text{highest}); w(\text{maximum}); w(\text{top}); w(\text{utmost})\}, \{w(\text{building}); w(\text{structure}); w(\text{construction})\}, \{w(\text{price}); w(\text{cost}); w(\text{fee}); w(\text{value})\})$

In effect, $w(t_i)$ is a function that calculates the weight for term $t_i$ in document $d$ using the proposed *BTF-IDF* term weighting scheme. In effect, the terms "maximum", "top", and "utmost" are synonyms for term "highest"; "structure" and "construction" are synonyms for term "building"; and "cost", "fee", and "value" are synonyms for term "price". Basically, all synonyms for a particular term $t_i$ are assigned a weight equal to the weight of the initial term $t_i$. Thus, the ultimate weighted *SWVM* vector for document $d$ can be given as follows:

$d = (\{3; 3; 3; 3\}, \{7; 7; 7\}, \{1; 1; 1; 1\})$

### 3.2. Generating Synonyms

Synonyms are different words but with similar meanings, for example, "car" and "automobile" are synonyms; and "buy" and "purchase" are synonyms too. The proposed *SWVM* considers the synonyms for every term it models. It utilizes a compiled thesaurus to assist in the generation of synonyms for every term during the indexing phase. A thesaurus is usually a list of words organized according to their similarities, differences, and other linguistic relationships. For every term $t_i$ in the vector $d$, a list of pertinent synonyms is generated $\{t_{i,1}; t_{i,2}; t_{i,3}; t_{i,s}\}$ where $i$ denotes a particular term in the vector and $s$ a particular synonym for the $i_{th}$ term. Then, every generated synonym is assigned a weight equal to the weight of the original term, that is $w(t_{i,s}) = w(t_i)$. The reason behind generating synonyms is to increase the precision of retrieval systems so that for instance a query containing the word "car" matches all documents containing the word "car", "automobile", "vehicle", and "motor".

### 3.3. The BTF-IDF Term Weighting Scheme

*BTF-IDF* is specifically designed to weight terms in a web document by exploiting the tag structure of its *HTML* language to deduce the magnitude of certain terms with regard to other terms and consequently to assign them higher weights. Put differently, it weights terms according to the tags in which they appear. For example, terms appearing in the *<title>* tag are considerably weighted as they might be strongly correlated to the semantics of the document. On the other hand, terms appearing in other tags such as *<td>*, *<p>*, and *<label>* do not receive extra credits as their semantic importance might be minimal. At heart, *BTF-IDF* is derived from the *TF-IDF* weighting scheme, however, altered with the aim of boosting the weight for terms located within some special *HTML* tags. The *BTF-IDF* scheme can be formally defined as:

$$w_{ij} = (tf_{ij} * boosting\_value) * idf_j$$

where $w_{ij}$ is the final weight for term $j$, $tf_{ij}$ is the frequency of term $j$ in document $i$, $idf_j$ is the Inverse Document Frequency of term $j$, and *boosting_value* is a constant number whose different values are outlined in Table 1.

**Table 1:** HTML tags and their boosting values

| HTML Tag | Boosting Value |
|---|---|
| <title> | 18 |
| <meta> | 16 |
| <h1> | 14 |
| Webpage URL | 18 |
| Other tags | 1 (No change) |

## 4. Experiments & Results

In the conducted experiments, a collection of 100 web documents pertaining to the IT field was compiled and deployed on a server computer. For comparison purposes, a sample document *d* (originally extracted from "www.microsoft.com/athome/setup/optimize.aspx") was selected out of the collection, then modeled and weighted using 1) the traditional *VSM* model and the *TF-IDF* scheme, and 2) the proposed *SWVM* model and the *BTF-IDF* scheme. Table 2 delineates the results obtained for the classical *TF-IDF* weighting scheme, ordered ascending according to their weight.

**Table 2:** Results Obtained for the TF-IDF Scheme

| Term | TF | IDF | TF*IDF |
|---|---|---|---|
| Computer | 52 | 0.008 | 0.41 |
| Optimize | 1 | 1.52 | 1.52 |
| Microsoft | 11 | 0.19 | 2.09 |
| Windows | 16 | 0.14 | 2.24 |
| Performance | 8 | 0.32 | 2.56 |
| Disk | 30 | 0.09 | 2.70 |
| Check | 26 | 0.12 | 3.12 |
| Error | 10 | 0.51 | 5.10 |
| … | | | |

Subsequently, the web document *d* can be represented using the *VSM* model as:

*d* = ( *w*(*computer*), *w*(*optimize*), *w*(*microsoft*), *w*(*windows*), *w*(*performance*), *w*(*disk*), *w*(*check*), *w*(*error*) )
*d* = ( *0.41*, *1.52*, *2.09*, *2.24*, *2.56*, *2.70*, *3.12*, *5.10* )

It is obvious that despite the web document *d* is concerned with "optimizing computer performance", the term "optimize" occurred only once in the document and was assigned a relatively low weight equal to 1.52. Likewise, the term "performance" was assigned a low weight compared to other terms. As a result, there is a low probability that a query such as "optimize computer performance" would return document *d*. Moreover, searching for other queries such as "improve computer speed" or "increase PC execution" would return no results as the *TF-IDF* does not take into account synonyms for terms in the document.

Using the proposed model, all terms weights were re-calculated, and synonyms were generated from a thesaurus for every term and were assigned the same weight as the original term. Table 3 outlines the results obtained for the *BTF-IDF* weighting scheme, ordered ascending according to their weight.

**Table 3:** Results obtained for the BTF-IDF scheme

| Term | Term Synonyms | Term Frequency | BTF | IDF | BTF*IDF |
|---|---|---|---|---|---|
| Computer | Workstation PC Processor | \<title\>: 1 \<h1\>: 1 Other: 51 | 1 * boosting_value(\<title\>) = 18 1 * boosting_value(\<h1\>) = 14 51 * boosting_value(other) = 51 **BTF = 18+14+51= 83** | 0.008 | 0.66 |
| Microsoft | Microsoft | Other: 11 | 11 * boosting_value(other) = 11 **BTF = 11** | 0.19 | 2.09 |
| Disk | Hard-disk Diskette | Other: 30 | 30 * boosting_value(other) = 30 **BTF = 30** | 0.09 | 2.70 |
| Check | Examine Test Try | Other: 26 | 26 * boosting_value(other) = 26 **BTF = 26** | 0.12 | 3.12 |
| Windows | Windows | \<meta\>: 1 Other: 16 | 1 * boosting_value(\<meta\>) = 16 16 * boosting_value(other) = 16 **BTF = 16+16 = 32** | 0.14 | 4.48 |
| Error | Fault Mistake Bug Glitch | Other: 10 | 10 * boosting_value(other) = 10 **BTF = 10** | 0.51 | 5.10 |
| Performance | Execution Efficiency Speed | \<meta\>: 1 \<h1\>: 1 Other: 7 | 1 * boosting_value(\<meta\>) = 16 1 * boosting_value(\<h1\>) = 14 7 * boosting_value(other) = 7 **BTF = 16+14+7= 37** | 0.32 | 11.84 |
| Optimize | Enhance Improve Boost | \<title\>: 1 \<meta\>: 1 \<h1\>: 1 URL: 1 Other: 0 | 1 * boosting_value(\<title\>) = 18 1 * boosting_value(\<meta\>) = 16 1 * boosting_value(\<h1\>) = 14 1 * boosting_value(URL) = 18 0 * boosting_value(other) = 0 **BTF = 18+16+14+18+0= 66** | 1.52 | 100.32 |

Evidently, weights for terms correlated with the document semantics such as "performance" and "optimize" were significantly boosted; while other terms less related to the semantics such as "windows" and "computer" were lightly boosted. In contrast, irrelevant terms such as "disk" and "check" retained their original weight and they were not boosted at all; however their significance with respect to the document was decreased. In sum, using *TF-IDF*, "performance" and "optimize" were among the terms with the lowest weight, whereas using *BTF-IDF*, they were among the terms with the highest weight. Eventually, document *d* can be represented using the proposed *SWVM* model as:

*d = ( { w(computer) ; w(workstation) ; w(PC) ; w(processor) } , { w(microsoft) } , { w(disk) ; w(hard-disk) ; w(diskette) } , { w(check) ; w(examine) ; w(try) ; w(test) } , { w(windows) } , { w(error) ; w(fault) ; w(mistake) ; w(bug) ; w(glitch) } , { w(performance) ; w(execution) ; w(efficiency) ; w(speed) } , { w(optimize) ; w(enhance) ; w(improve) ; w(boost) } )*

*d = ( { 0.66 ; 0.66 ; 0.66 ; 0.66 } , { 2.09 } , { 2.70 ; 2.70 ; 2.70 } , { 3.12 ; 3.12 ; 3.12 ; 3.12 } , { 4.48 } , { 5.10 ; 5.10 ; 5.10 ; 5.10 ; 5.10 } , { 11.84 ; 11.84 ; 11.84 ; 11.84 } , { 100.32 ; 100.32 ; 100.32 ; 100.32 } )*

For further experiments, a query *q*="optimize computer performance" was executed on both models. It is composed of three terms closely related to the semantics of document *d*, and it is represented using *SWVM* as:

*q = ( w(optimize), w(computer), w(performance) )*
*q = ( 1, 1, 1 )*

In order to evaluate the retrieval results, the similarity between the two vectors *d* and *q* was computed using the cosine metric where the normalization of the two vectors is integrated directly into the similarity equation. The complete math is sketched as follows:

Using the traditional *VSM* and the *TF-IDF*:

$$|\vec{d_j}| = \sqrt{\sum_{i=1}^{N} w_{i,j}^2} = \sqrt{0.41^2 + 1.52^2 + 2.09^2 + 2.24^2 + 2.56^2 + 2.70^2 + 3.12^2 + 5.10^2}$$

$$= \sqrt{61.45} = 7.83$$

$$|\vec{q}| = \sqrt{\sum_{i=1}^{N} w_{i,q}^2} = \sqrt{1^2 + 1^2 + 1^2}$$

$$= \sqrt{3} = 1.73$$

$$\vec{d_j} \bullet \vec{q} = 1.52*1 + 0.41*1 + 2.56*1 = 4.48$$

$$\frac{\vec{d_j} \bullet \vec{q}}{|\vec{d_j}||\vec{q}|} = 4.48 / 7.83 * 1.73 = 0.33$$

Using the proposed *SWVM* and the *BTF-IDF*:

$$|\vec{d_j}| = \sqrt{\sum_{i=1}^{N} w_{i,j}^2} = \sqrt{0.66^2 + 2.09^2 + 2.70^2 + 3.12^2 + 4.48^2 + 5.10^2 + 11.84^2 + 100.32^2}$$

$$= \sqrt{10{,}272.19} = 101.35$$

$$|\vec{q}| = \sqrt{\sum_{i=1}^{N} w_{i,q}^2} = \sqrt{1^2 + 1^2 + 1^2}$$

$$= \sqrt{3} = 1.73$$

$$\vec{d_j} \bullet \vec{q} = 100.32*1 + 0.66*1 + 11.84*1 = 112.82$$

$$\frac{\vec{d_j} \bullet \vec{q}}{|\vec{d_j}||\vec{q}|} = 112.82 / 101.35 * 1.73 = 0.64$$

The above results clearly show that using the standard *VSM* and *TF-IDF* produced a cosine value of 0.33, and thus a weak match exists between $q$ and $d$. Nevertheless, using the proposed *SWVM* and *BTF-IDF* yielded to a higher value equal to 0.64, and thus there is a higher probability that query $q$ will return document $d$. (the more the cosine value is close to 1, the more are the two vectors similar).

Moreover, another query $q$="improve PC speed" was executed, which contains terms not actually present in document $d$, but are synonyms to terms in document $d$. Query $q$ can be represented using *SWVM* as:

$q = ( w(improve), w(PC), w(speed) )$
$q = ( 1, 1, 1 )$

Calculating the cosine similarity between the new query $q$ and the original document $d$, yielded to the below results:

Using the traditional *VSM* and the *TF-IDF*:

$$|\vec{d_j}| = \sqrt{\sum_{i=1}^{N} w_{i,j}^2} = \sqrt{0.41^2 + 1.52^2 + 2.09^2 + 2.24^2 + 2.56^2 + 2.70^2 + 3.12^2 + 5.10^2}$$

$$= \sqrt{61.45} = 7.83$$

$$|\vec{q}| = \sqrt{\sum_{i=1}^{N} w_{i,q}^2} = \sqrt{1^2 + 1^2 + 1^2}$$

$$= \sqrt{3} = 1.73$$

$$\vec{d_j} \bullet \vec{q} = 0 \text{ (share no common terms)}$$

$$\frac{\vec{d_j} \bullet \vec{q}}{|\vec{d_j}||\vec{q}|} = 0 / 7.83 * 1.73 = 0$$

Using the proposed *SWVM* and the *BTF-IDF*:

$$|\vec{d_j}| = \sqrt{\sum_{i=1}^{N} w_{i,j}^2} = \sqrt{0.66^2 + 2.09^2 + 2.70^2 + 3.12^2 + 4.48^2 + 5.10^2 + 11.84^2 + 100.32^2}$$

$$= \sqrt{10{,}272.19} = 101.35$$

$$|\vec{q}| = \sqrt{\sum_{i=1}^{N} w_{i,q}^2} = \sqrt{1^2 + 1^2 + 1^2}$$

$$= \sqrt{3} = 1.73$$

$$\vec{d_j} \bullet \vec{q} = 100.32*1 + 0.66*1 + 11.84*1 = 112.82$$

$$\frac{\vec{d_j} \bullet \vec{q}}{|\vec{d_j}||\vec{q}|} = 112.82 / 101.35 * 1.73 = 0.64$$

Using *TF-IDF*, no match was observed between $d$ and $q$ as the similarity equation resulted in a cosine value of 0; this means that both vectors share no common terms. Alternatively, using the *BTF-IDF*, a higher cosine value of 0.64 was obtained, and that is due to the advantage of the *SWVM* model that generates synonyms for terms in the document and assigns them weights equal to the weights of their corresponding original terms. In effect, query $q$ and document $d$ do not share literally the same vocabulary terms, however, they do share synonym terms.

## 5. Conclusions and Future Work

Classical vector space models such as *VSM* and term weighting schemes such as *TF* and *TF-IDF* were not initially designed to support different formats of text documents as they treat their enclosed terms uniformly. This paper proposed a novel semantic-sensitive web vector model called *SWVM* and a term weighting scheme called *BTF-IDF* intended for web *IR* systems. Principally, the proposed model exploits the *HTML* tag structure of web documents to deduce the semantic importance of specific terms with respect to others. Consequently, terms that appear in certain pre-specified tags are assigned a higher weight. Additionally, synonyms for every single term in the document are generated, then assigned a weight equal to their corresponding terms, and stored as extra features in the vector model. As by specifications, the *HTML* language features special-type tags to emphasize any content that is subject to great search hits, so should be the web *IR* model able to favor more weighty content over less weighty content in order to reflect the original design of the document. Likewise, synonyms are important to model as users who search for specific content tend to express their intentions with different vocabulary words. As a result, when experimented, the proposed model clearly revealed a remarkable retrieval precision as more relevant documents were successfully retrieved despite the different wordings in the input query.

Further research can improve upon the proposed model so much so that it supports the retrieval of non-textual content in multimedia-rich web documents such as images and videos. In addition, more investigations are to be carried out to extend the proposed model so that it can be used to categorize and sort web documents according to their topics, subjects, contents, and semantics.

## Acknowledgments

This research was funded by the Lebanese Association for Computational Sciences (LACSC), Beirut, Lebanon under the "Web Information Retrieval Research Project – WIRRP2011".